\documentclass{article}
\usepackage{spconf,amsmath,graphicx}
\usepackage{hyperref} 
\title{Msdtron: a high-capability multi-speaker speech synthesis system for diverse data using characteristic information}
\name{Qinghua Wu \qquad Quanbo Shen \qquad Jian Luan \qquad Yujun Wang}
\address{Xiaomi Corporation, Beijing, China}

\begin{document}
\topmargin=0mm
%
\maketitle
\begin{abstract}
In multi-speaker speech synthesis, data from a number of speakers usually tend to have great diversity due to the fact that the speakers may differ largely in ages, speaking styles, emotions, and so on. It is important but challenging to improve the modeling capabilities for multi-speaker speech synthesis. To address the issue, this paper proposes a high-capability speech synthesis system, called Msdtron, in which 1) a representation of the harmonic structure of speech, called excitation spectrogram, is designed to directly guide the learning of harmonics in mel-spectrogram. 2) conditional gated LSTM (CGLSTM) is proposed to control the flow of text content information through the network by re-weighting the gates of LSTM using speaker information. The experiments show a significant reduction in reconstruction error of mel-spectrogram in the training of the multi-speaker model, and a great improvement is observed in the subjective evaluation of speaker adapted model.
\end{abstract}
\begin{keywords}
multi-speaker speech synthesis, voice cloning, speaker adaptation, few-shot speech synthesis
\end{keywords}
\section{Introduction}
\label{sec:intro}

Neural text-to-speech is very popular in recent years \cite{Ren2020FastSpeech2F,Shen2018NaturalTS}, and it can already produce speech that's almost as natural as speech from a real person with high voice quality. However, data collection is still a big challenge. We often need to collect a large amount of data in a high-fidelity recording studio with professional guidance to obtain high voice quality and high consistency of recordings. It is very costly, time-consuming, or even impossible, e.g. in cases of custom speech and Lombard speech \cite{Bollepalli2019LombardSS}. Meanwhile, noisy and diverse data is usually easier to be collected. Thereby multi-speaker speech synthesis is proposed, which uses diverse data from lots of speakers to train a robust multi-speaker generative model. It can be further adapted in different tasks such as speaker adaptation \cite{Hu2019NeuralTA}, cross-lingual text-to-speech synthesis \cite{Chen2019CrossLingualMT}, and style conversion \cite{Paul2020EnhancingSI}. 

The state-of-art systems have an encoder-decoder structure network with speaker embedding as additional inputs \cite{Paul2020EnhancingSI, Choi2020AttentronFT, Chien2021InvestigatingOI, Cooper2020ZeroShotMT, Jia2018TransferLF}. Some works investigated the effective representations of speakers, e.g. \cite{Chien2021InvestigatingOI, Cooper2020ZeroShotMT} studied the effects of different speaker embeddings such as x-vector \cite{Snyder2018XVectorsRD}, LDE-based speaker encoding \cite{Cooper2020ZeroShotMT}. \cite{Choi2020AttentronFT} proposed an attention-based variable-length embedding. \cite{Cai2020FromSV} measured the speaker similarity between the predicted mel-spectrogram and the reference. Some works focused on solving the problem of noisy data \cite{Hu2019NeuralTA, Paul2020EnhancingSI}, e.g. \cite{Paul2020EnhancingSI} did research into the methods of transfer learning for noisy samples. \cite{Hsu2019DisentanglingCS} aimed to disentangle speaker embedding and noise by data augmentation and a conditional generative model. 
And some works were interested in the controllability of systems in the manner of zero-shot. \cite{Choi2020AttentronFT, Cooper2020ZeroShotMT} tried to obtain target voice by feeding target speaker embedding without speaker adaptation. \cite{Hsu2019HierarchicalGM} introduced latent variables to control the speaking style.

The previous studies rarely gave insights into what role the characteristic information such as timbre and style played. The characteristic information is usually represented by a fixed-or-variable-length embedding which may not be effective enough, e.g. the pitch embedding is relevant to the harmonics of speech but it's not an effective representation of the harmonic structure. Besides, the embedding of characteristic information is typically concatenated or added to the text content representation or is simply used to perform an affine transformation on it \cite{Kumar2020FewSA}. In this way, the characteristic information is playing a similar role as the text content information in the network, which is not what we expected. 

In this paper, we propose an encoder-decoder structured speech synthesis system (Msdtron), investigating the effective use of characteristic information. The major contributions are: 1) Excitation spectrogram is designed to explicitly characterize the harmonic structure of speech. It acts as a skeleton of the target mel-spectrogram to optimize the learning process. 2) Conditional gated LSTM (CGLSTM) is proposed whose input/output/forget gates are re-weighted by speaker embedding while the cell/hidden states are dependent on the text content. That’s to say, the speaker embedding controls the flow of text content information by gates without affecting cell state and hidden state directly.

The rest article is organized as follows: Section 2 describes the framework of the proposed system (Section 2.1), excitation spectrogram generator (Section 2.2), and CGLSTM in the decoder (Section 2.3). Section 3 is about the detailed settings and results of experiments. Finally, a conclusion is drawn in Section 4. 

\section{Msdtron}
\label{sec:mspkm}

\subsection{Framework}

\begin{figure}[t]
  \centering
  \includegraphics[scale=1.0]{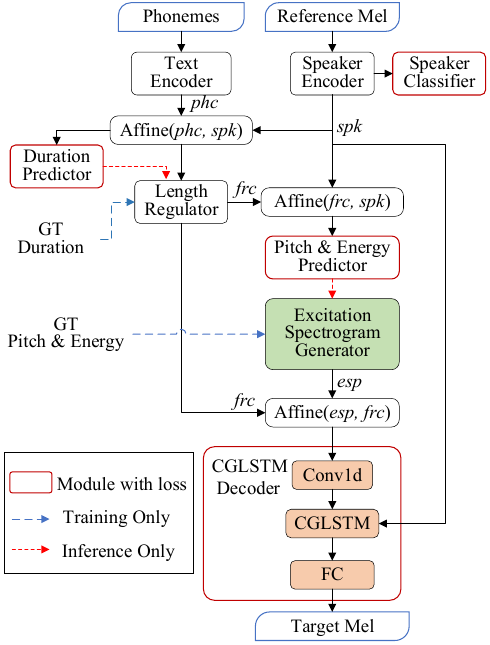}
  \caption{The overall framework of Msdtron. (p.s. GT means ground-truth)}
  \label{fig:framework}
\end{figure}

The framework of Msdtron is illustrated as Figure~\ref{fig:framework}, a state-of-art encoder-decoder structure.
The text-encoder is a standard Tacotron2 encoder \cite{Shen2018NaturalTS} which has a stacked Conv1d followed by BLSTM. It takes phoneme sequences as inputs and learns an embedding of the text content. Besides, a sentence-level speaker embedding is extracted from the reference mel-spectrogram by a GST-like \cite{Wang2018StyleTU} speaker-encoder, with a stacked Conv2d+BatchNorm reference encoder, and a multi-head attention \cite{Vaswani2017AttentionIA}. Instead of introducing Gradient Reversal Layer (GRL) \cite{Ganin2015UnsupervisedDA} to remove the text content information from the speaker embedding, the reference is randomly chosen from the utterances of the same speaker with the target \cite{Jia2018TransferLF}. 

Similar to \cite{Ren2020FastSpeech2F}, a length regulator is used for the alignment between phoneme sequences and target mel-spectrogram sequences. The duration predictor is simply a layer of BLSTM with pre-dense and post-dense layers.

Pitch and energy are predicted separately with the same network structure, a stacked Conv1d with post-dense. Then excitation spectrogram is determined by the pitch and energy (see Section 2.2), which explicitly represents the harmonic structure of speech.

The decoder is to predict the target mel-spectrogram with the proposed CGLSTM (see Section 2.3) in it.

Finally, in the flow of information, affine transformations are carried out. It is defined as follows:

\begin{equation}
  affine(x_{i}, x_{c})=x_{i}\odot proj^{1}(x_{c})+proj^{2}(x_{c})
  \label{eq:affine}
\end{equation}
\begin{equation}
  proj(x)=w\otimes x+b
  \label{eq:proj}
\end{equation}
where $\odot$ means element-wise multiplication, and $\otimes$ means matrix multiplication.

\subsection{Excitation Spectrogram Generator}
In the source-filter analysis \cite{Morise2016WORLDAV}, speech is produced when an excitation signal passes through a filter representing the vocal track characteristics. It's composed of harmonic and noise components, as a result of periodic or aperiodic signals modulated by the vocal track filter. Existing studies usually use pitch information to model the periodicity of speech. Unfortunately, the pitch has a poor ability to express the harmonic characteristics of speech. Thus an explicit representation of harmonic structure in the spectrogram is proposed, called excitation spectrogram, in this session.

To make things easy, it is assumed that speech signals consist of harmonic components only at voiced segments, and noise components only at unvoiced segments. The production of voiced speech is modeled as a time-varying periodic impulse passing through the vocal track filter and generating harmonics at multiples of the periodic frequency as Equation~\ref{eq:har}. The unvoiced speech is generated from a white Gaussian noise signal. 

\begin{equation}
  har_{i} = i \ast f_{0} \:\:\:\:\: i\in [1, N_{h}] 
  \label{eq:har}
\end{equation}
where $har_{i}$ means the $i^{th}$ harmonic frequency of speech, $N_{h}$ is the number of harmonics, and $f0$ is the periodic (fundamental) frequency.

In the excitation spectrogram, energy is supposed evenly distributed on each harmonic frequency of voiced segments or the whole frequency band of unvoiced segments. It can help to learn the accurate distribution of energy in the frequency of mel-spectrogram. Thus, the excitation spectrogram can be defined as Equation~\ref{eq:els}.

\begin{equation}
  els_{i, j} = \left\{\begin{matrix}
      e_{i}/N_{f}\; \; \; & if \; i\in unvoiced\; \; \; \; \; \;\; \; \; \; \;\; \; \; \; \; \; \; \; \; \;  \\ 
      e_{i}/N_{h}\; \; \; & if \; i\in voiced,\:j\in harmonics\\ 
      0\; \; \; \; \; \;\; \; \; &  if \; i\in voiced,\:j\notin harmonics
\end{matrix}\right. 
  \label{eq:els}
\end{equation}

where $els_{i, j}$ means the $j^{th}$ linear spectrum at $i^{th}$ frame, $e_{i}$ is the total energy of $i^{th}$ frame, and $N_{f}$ is the fft number in the calculation of linear spectrum.

Finally, the linear excitation spectrogram $els$ is converted to mel excitation spectrogram $esp$ by Equation~\ref{eq:esp}

\begin{equation}
  esp = els \otimes W^{l2m}_{N_{f} \times N_{m}} 
  \label{eq:esp}
\end{equation}

where $W^{l2m}_{N_{f} \times N_{m}}$ is the transformation matrix from linear to mel spectrogram and $N_{m}$ is the dimension of mel-spectrogram.

\subsection{Conditional Gated LSTM}

Text content is the most important feature of speech due to its decisive role in intelligibility. In addition, speech can be characterized in terms of timbre, style, speed, and emotion, etc. Many researches aim to change or control some of these characteristics without a negative influence on the intelligibility and voice quality. For this purpose, the embedding of characteristic information is usually added or concatenated to the text content embedding, which is fed as the inputs of network. However, in this way, the characteristic information plays a similar role to the text content. Both of them would directly take effect in the same way on the intelligibility, voice quality, and other characteristics at the same time, which is not the way we expected. Consequently, we propose conditional gated LSTM (CGLSTM) where the characteristic information is used to re-weight the gates and the text content flows in the hidden/cell states. Thereby the characteristic information will directly play a part in the gates-based flow of the text content without operating the text content in itself.

Compared with LSTM, which is frequently used in speech synthesis tasks due to its good capacity of learning long dependencies, CGLSTM uses the characteristic information to re-weight the input/output/forget gates as Equation~\ref{eq:gate}.

\begin{equation}
  g_{t} = \sigma(( W_{xg} \otimes [h_{t-1}, x_{t}] + b_{xg} ) \odot ( W_{cg} \otimes c_{t} + b_{cg} ))
  \label{eq:gate}
\end{equation}

where $g_{t}$ can be the forget, input or output gate; $x_{t}$, $c_{t}$, and $h_{t-1}$ are the current text content inputs, current characteristic information embedding, and previous hidden state; $W$ and $b$ are the corresponding weights and biases.

\section{Experiments}
\label{sec:exp}

\subsection{Corpus}
The data set of our experiments is the public multi-speaker mandarin corpus AISHELL-3 \cite{Shi2020AISHELL3AM}, which contains roughly 85 hours of recordings spoken by 218 native Chinese mandarin speakers. Among them, recordings from 173 speakers have Chinese character-level and pinyin-level transcripts and a total of 63263 utterances. This part of the transcribed data will be used in our experiments, which is divided into the training set and test set without overlapping.

\begin{itemize}
\item \textbf{Training set}: contains 57304 utterances from 165 speakers, with 133 females 46915 utterances and 32 males 10389 utterances. The training set is used to pre-train the multi-speaker generative model, which is further adapted using the test set. 

\item \textbf{Test set}: contains 4 females and 4 males, and only 20 utterances of each speaker are randomly chosen for speaker adaptation.
\end{itemize}

The recordings are mono, 16bit, and down-sampled from 44100HZ to 16000HZ. Preprocessing is conducted on both the training and the test sets to reduce the diversity of them: 1) Energy normalization by scaling the maximal amplitude of utterance. 2) Silence trimming by keeping 60ms silence at the head and tail of utterance.

\subsection{Setup}
The pipeline of our experiments includes 1) Pre-training: train the multi-speaker generative model using the training set. 2) Speaker adaptation: train the target model by transfer learning using the single-speaker data from the test set and 3) Synthesis: infer the mel-spectrogram and synthesize the waveform by Hifi-Gan vocoder\cite{kong2020hifi}.

In our experiments, the frame hop size is set to 12.5ms, the window size 50ms, and the number of mel-bank 80 for mel-spectrogram. Mean Absolute Error (MAE) is calculated to measure the reconstruction error of pitch and energy while Mean Square Error (MSE) is applied to mel-spectrogram. Besides, the task of speaker classification uses cross-entropy as the loss function. The setup of our experiments is described as follows:

\begin{itemize}

\item \textbf{Baseline}: A modified Tacotron2 with a length regulator replacing attention-based alignment. In order to cope with the case of multi-speakers, a speaker-encoder is added. In detail, following modifications are made on the proposed system (Figure~\ref{fig:framework}): 1) The excitation spectrogram generator is removed. 2) CGLSTM in the decoder is replaced with the standard LSTM while an affine transformation is conducted on the speaker embedding and the text content before they are fed to the decoder.

\item \textbf{System-1}: Baseline + excitation spectrogram generator, which is the system Msdtron without CGLSTM-decoder.

\item \textbf{System-2}: Baseline + excitation spectrogram generator + CGLSTM decoder. It is our proposed system Msdtron in Figure~\ref{fig:framework}.

\end{itemize}

\subsection{Multi-speaker Model}
\begin{figure}[th]
  \centering
  \includegraphics[scale=1.0]{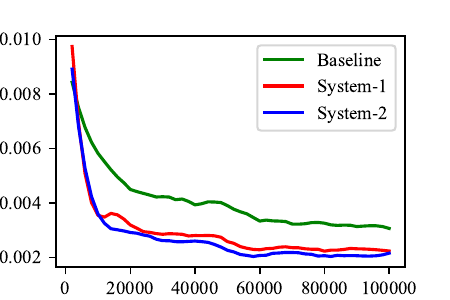}
  \caption{Reconstruction Errors of mel-spectrogram in different systems. (x-label: steps of training; y-label: mel-error) }
  \label{fig:mel_loss}
\end{figure}

Figure~\ref{fig:mel_loss} shows the reconstruction error of mel-spectrogram of different systems in the pre-training stage. Compared with the Baseline, the excitation-spectrogram generator brought an obvious improvement in terms of reconstruction error in System-1. The reconstruction error reduced further in System-2 when the CGLSTM-decoder was introduced. It indicates that the excitation spectrogram and CGLSTM can greatly improve the modeling capability of systems for multi-speaker corpus.

We also compared the number of parameters of systems in Table~\ref{tab:para}. It drops by 10\% from Baseline to System-2. In other words, we can achieve better performance with less computation and less memory.

\begin{table}[th]
\caption{The number of parameters of systems}
\label{tab:para}
\begin{center}
\begin{tabular}{ccc}
\hline
\textbf{Baseline}& \textbf{System-1}& \textbf{System-2} \\ \hline
11.24 M          & 9.5 M            & 10.08 M           \\ \hline
\end{tabular}
\end{center}
\end{table}

\subsection{Speaker Adapted Model}
For unseen speakers in the test set, we adapted the multi-speaker model using data of the target speaker\footnote[1]{Audio examples: \url{https://Msdtron.github.io/}} . The Mean Opinion Score (MOS) test was carried out to evaluate its performance in voice quality of speech and speaker similarity. 9 native Chinese testers participated in it. The MOS results are shown in Table~\ref{tab:ft_qs}.

\begin{table}[th]
\caption{MOS of voice quality (Quality) and speaker similarity (Similarity) for unseen-speaker after speaker adaptation. In the evaluations, scores range from 1 to 5. 1) For voice quality, score=1 means the voice has strong and annoying noise, or bad pronunciation while score=5 means the voice is clean, pleasant, and clear pronounced. 2) For speaker similarity, score=1 means the compared two voices don't sound like from the same person at all while score=5 means it's easy to make a judgment that they are from the same person.}
\label{tab:ft_qs}
\begin{center}
\begin{tabular}{lllll}
\hline
                & \multicolumn{2}{c}{\textbf{Quality}} & \multicolumn{2}{c}{\textbf{Similarity}} \\ \cline{2-5} 
                & female        & male          & female          & male          \\ \hline
Ground-truth    & 4.70          & 4.42          & -               & -             \\ \hline
Baseline        & 2.71          & 2.76          & 3.39            & 3.28          \\ \hline
System-1        & 3.67          & 3.06          & 3.89            & 3.31          \\ \hline
System-2        & \textbf{3.97} & \textbf{3.58} & \textbf{4.06}   & \textbf{3.64} \\ \hline
\end{tabular}
\end{center}
\end{table}

According to the MOS results, System-1 outperformed the Baseline in both aspects of voice quality and speaker similarity. It indicates that the excitation spectrogram is much more effective than the simple use of pitch and energy. It can reduce the noise or signal distortion caused by insufficient modeling capabilities for complex data, thus improving the clearness of pronunciation. Besides, System-2 achieved the best performance, which proves that CGLSTM can control the specific characteristics of voice better than LSTM while the voice quality is improved at the same time.

Furthermore, more improvement was observed on females by comparing Baseline and System-1 while more was observed on males from System-1 to System-2. One possible reason is the unbalanced data for females and males in the training set, with a rough ratio female : male = 9 : 2. When the excitation spectrogram was used, it helped System-1 to learn more information of females than of males and thus to achieve better performance on females. Meanwhile, in System-2 with CGLSTM-decoder, the speaker embedding is attempted to control the characteristics of speakers without a negative impact on voice quality. Therefore, System-2 could share knowledge better apart from speaker information between males and females, which brought obvious improvement on males.

Finally, we notice that the voice quality of Baseline is relatively low. The main flaw of the synthesized voices is the annoying noise. It can be partly explained by the fact that the corpus AISHELL-3 contains a certain reverberation and background noise, which often degrades the performance of TTS systems. We can try to add additional professional recordings into the training set, just as the M2VoC-2021 challenge did. It also proves that Tacotron2 and its simple variants don't perform well on diverse multi-speaker data.  

\section{Conclusions}
\label{sec:conclu}

In this paper, we have proposed the excitation spectrogram to represent the harmonic structure of speech, and CGLSTM to better control the specific characteristics of speech with less impact on the voice quality than LSTM. The proposed system (Msdtron) outperformed the baseline largely both on the voice quality and the speaker similarity. More researches will be conducted to enhance the cleanliness and expressiveness of synthesized voices with more diverse data of different noise levels and speaking styles in the future.

\vfill\pagebreak


\bibliographystyle{IEEEbib}
\bibliography{refs}

\begin{thebibliography}{10}

\bibitem{Ren2020FastSpeech2F}
Yi~Ren, C.~Hu, Xu~Tan, Tao Qin, Sheng Zhao, Zhou Zhao, and T.~Liu,
\newblock ``Fastspeech 2: Fast and high-quality end-to-end text to speech,''
\newblock {\em ArXiv}, vol. abs/2006.04558, 2020.

\bibitem{Shen2018NaturalTS}
Jonathan Shen, R.~Pang, Ron~J. Weiss, M.~Schuster, Navdeep Jaitly, Z.~Yang,
  Z.~Chen, Yu~Zhang, Yuxuan Wang, R.~Skerry-Ryan, R.~A. Saurous, Yannis
  Agiomyrgiannakis, and Y.~Wu,
\newblock ``Natural tts synthesis by conditioning wavenet on mel spectrogram
  predictions,''
\newblock {\em 2018 IEEE International Conference on Acoustics, Speech and
  Signal Processing (ICASSP)}, pp. 4779--4783, 2018.

\bibitem{Bollepalli2019LombardSS}
Bajibabu Bollepalli, Lauri Juvela, and P.~Alku,
\newblock ``Lombard speech synthesis using transfer learning in a tacotron
  text-to-speech system,''
\newblock in {\em INTERSPEECH}, 2019.

\bibitem{Hu2019NeuralTA}
Qiong Hu, E.~Marchi, David Winarsky, Y.~Stylianou, Devang~K. Naik, and
  Sachin~S. Kajarekar,
\newblock ``Neural text-to-speech adaptation from low quality public
  recordings,''
\newblock in {\em Speech Synthesis Workshop 10}, 2019.

\bibitem{Chen2019CrossLingualMT}
Mengnan Chen, Minchuan Chen, S.~Liang, J.~Ma, Lei Chen, Shaojun Wang, and
  J.~Xiao,
\newblock ``Cross-lingual, multi-speaker text-to-speech synthesis using neural
  speaker embedding,''
\newblock in {\em INTERSPEECH}, 2019.

\bibitem{Paul2020EnhancingSI}
D.~Paul, P.~V.~M. Shifas, Yannis Pantazis, and Y.~Stylianou,
\newblock ``Enhancing speech intelligibility in text-to-speech synthesis using
  speaking style conversion,''
\newblock {\em ArXiv}, vol. abs/2008.05809, 2020.

\bibitem{Choi2020AttentronFT}
Seungwoo Choi, Seungju Han, Dong-Young Kim, and Sungjoo Ha,
\newblock ``Attentron: Few-shot text-to-speech utilizing attention-based
  variable-length embedding,''
\newblock {\em ArXiv}, vol. abs/2005.08484, 2020.

\bibitem{Chien2021InvestigatingOI}
Chung-Ming Chien, Jheng hao Lin, Chien yu~Huang, Po~chun Hsu, and Hung yi~Lee,
\newblock ``Investigating on incorporating pretrained and learnable speaker
  representations for multi-speaker multi-style text-to-speech,''
\newblock {\em ArXiv}, vol. abs/2103.04088, 2021.

\bibitem{Cooper2020ZeroShotMT}
E.~Cooper, Cheng-I Lai, Y.~Yasuda, Fuming Fang, Xin~Eric Wang, Nanxin Chen, and
  J.~Yamagishi,
\newblock ``Zero-shot multi-speaker text-to-speech with state-of-the-art neural
  speaker embeddings,''
\newblock {\em ICASSP 2020 - 2020 IEEE International Conference on Acoustics,
  Speech and Signal Processing (ICASSP)}, pp. 6184--6188, 2020.

\bibitem{Jia2018TransferLF}
Ye~Jia, Y.~Zhang, Ron~J. Weiss, Q.~Wang, Jonathan Shen, Fei Ren, Z.~Chen,
  P.~Nguyen, R.~Pang, I.~Lopez-Moreno, and Y.~Wu,
\newblock ``Transfer learning from speaker verification to multispeaker
  text-to-speech synthesis,''
\newblock in {\em NeurIPS}, 2018.

\bibitem{Snyder2018XVectorsRD}
David Snyder, D.~Garcia-Romero, G.~Sell, D.~Povey, and S.~Khudanpur,
\newblock ``X-vectors: Robust dnn embeddings for speaker recognition,''
\newblock {\em 2018 IEEE International Conference on Acoustics, Speech and
  Signal Processing (ICASSP)}, pp. 5329--5333, 2018.

\bibitem{Cai2020FromSV}
Zexin Cai, C.~Zhang, and Ming Li,
\newblock ``From speaker verification to multispeaker speech synthesis, deep
  transfer with feedback constraint,''
\newblock {\em ArXiv}, vol. abs/2005.04587, 2020.

\bibitem{Hsu2019DisentanglingCS}
Wei-Ning Hsu, Yu~Zhang, Ron~J. Weiss, Yu-An Chung, Yuxuan Wang, Y.~Wu, and
  James~R. Glass,
\newblock ``Disentangling correlated speaker and noise for speech synthesis via
  data augmentation and adversarial factorization,''
\newblock {\em ICASSP 2019 - 2019 IEEE International Conference on Acoustics,
  Speech and Signal Processing (ICASSP)}, pp. 5901--5905, 2019.

\bibitem{Hsu2019HierarchicalGM}
Wei-Ning Hsu, Y.~Zhang, Ron~J. Weiss, H.~Zen, Y.~Wu, Yuxuan Wang, Yuan Cao,
  Y.~Jia, Z.~Chen, Jonathan Shen, P.~Nguyen, and R.~Pang,
\newblock ``Hierarchical generative modeling for controllable speech
  synthesis,''
\newblock {\em ArXiv}, vol. abs/1810.07217, 2019.

\bibitem{Kumar2020FewSA}
N.~Kumar, Srishti Goel, A.~Narang, and Brejesh Lall,
\newblock ``Few shot adaptive normalization driven multi-speaker speech
  synthesis,''
\newblock {\em ArXiv}, vol. abs/2012.07252, 2020.

\bibitem{Wang2018StyleTU}
Yuxuan Wang, Daisy Stanton, Yu~Zhang, R.~Skerry-Ryan, Eric Battenberg, Joel
  Shor, Y.~Xiao, Fei Ren, Ye~Jia, and R.~A. Saurous,
\newblock ``Style tokens: Unsupervised style modeling, control and transfer in
  end-to-end speech synthesis,''
\newblock in {\em ICML}, 2018.

\bibitem{Vaswani2017AttentionIA}
Ashish Vaswani, Noam Shazeer, Niki Parmar, Jakob Uszkoreit, Llion Jones,
  Aidan~N. Gomez, L.~Kaiser, and Illia Polosukhin,
\newblock ``Attention is all you need,''
\newblock {\em ArXiv}, vol. abs/1706.03762, 2017.

\bibitem{Ganin2015UnsupervisedDA}
Yaroslav Ganin and V.~Lempitsky,
\newblock ``Unsupervised domain adaptation by backpropagation,''
\newblock {\em ArXiv}, vol. abs/1409.7495, 2015.

\bibitem{Morise2016WORLDAV}
Masanori Morise, Fumiya Yokomori, and K.~Ozawa,
\newblock ``World: A vocoder-based high-quality speech synthesis system for
  real-time applications,''
\newblock {\em IEICE Trans. Inf. Syst.}, vol. 99-D, pp. 1877--1884, 2016.

\bibitem{Shi2020AISHELL3AM}
Yao Shi, Hui Bu, Xin Xu, Shaoji Zhang, and Ming Li,
\newblock ``Aishell-3: A multi-speaker mandarin tts corpus and the baselines,''
\newblock {\em ArXiv}, vol. abs/2010.11567, 2020.

\bibitem{kong2020hifi}
Jungil Kong, Jaehyeon Kim, and Jaekyoung Bae,
\newblock ``Hifi-gan: Generative adversarial networks for efficient and high
  fidelity speech synthesis,''
\newblock {\em arXiv preprint arXiv:2010.05646}, 2020.

\end{thebibliography}

\end{document}